\documentclass[namedreferences]{solarphysics}

\usepackage[hyperref,optionalrh,showbiblabels]{spr-sola-addons} % For Solar Physics 
\usepackage{graphicx}        % For eps figures, newer & more powerfull
\usepackage{color}           % For color text: \color command
\usepackage{breakurl}        % For breaking URLs easily trough lines
\newcommand{\adv}{    {\it Adv. Space Res.}} 
 
\newcommand{\aap}{    {\it Astron. Astrophys.}}

\newcommand{\aapr}{   {\it Astron. Astrophys. Rev.}}

\newcommand{\apj}{    {\it Astrophys. J.}}

\newcommand{\apss}{   {\it Astrophys. Space Sci.}}

\newcommand{\solphys}{{\it Solar Phys.}}
 
\newcommand{\ssr}{    {\it Space Sci. Rev.}} 
\chardef\us=`\_

\begin{document}

\begin{article}
\begin{opening}

\title{On the nature of the magnetic field asymmetry in magnetically coupled leading and following sunspots observed in active regions with no eruptive events\\ {\it Solar Physics}}

\author{Iu.S.~\surname{Zagainova}$^{1}$\sep
        V.G.~\surname{Fainshtein}$^{2}$\sep
        G.V.~\surname{Rudenko}$^{2}$\sep
        V.N.~\surname{Obridko}$^{1}$}

   \institute{$^{1}$ IZMIRAN (Institute of Earth magnetism, ionosphere and radiowaves propagation named after Nikolay Pushkov of the Russian Academy of Sciences), 142190, Moscow, Troitsk, Russia. email: \url{yuliazag@izmiran.ru} email: \url{obridko@izmiran.ru}\\                     
              $^{2}$ ISTP SB RAS (Institute of Solar-Terrestrial Physics of Siberian Branch of the Russian Academy of Sciences), 664033, Irkutsk, P/O Box 291, Russia. email: \url{vfain@iszf.irk.ru} \\
             }

\runningauthor{Zagainova et al.}
\runningtitle{Leading and following sunspots}

\begin{abstract}
In this study, we investigate magnetic properties of umbra of magneto-conjugate leading and following sunspots, i.e. connected through magnetic field lines. We established dependences between individual sunspot umbra field characteristics, and between these characteristics on the umbra area ($S$) separately for sunspot pairs, for which the minimal angle between the umbra magnetic field line of the leading ($L$) sunspot and the positive normal line to the Sun surface is smaller, than that in the following ($F$) sunspot ($\alpha_{min-L}<\alpha_{min-F}$; such sunspot pairs are the bulk) and, on the contrary, when $\alpha_{min-L}>\alpha_{min-F}$. The $\alpha_{min-L}(S_L)$, $\alpha_{min-F}(S_F)$, $B_{max-L}(S_L)$ and $B_{max-F}(S_F)$ dependences are shown to have similar behavior features, and are quantitatively close for two sunspot groups with a different asymmetry of the sunspot magnetic field connecting them (here, $B_{max-L,F}(S_L)$ is the magnetic induction maximum induction in umbrae of the leading and the following sunspots). The dependence of mean values of angles within umbra $<\alpha_{L,F}>$ on the sunspot umbra area $S_{L,F}$ and on the mean value of magnetic induction in umbra $<B_{L,F}>$ appeared different for two cases. Also, in the bulk of the investigated sunspot pairs, the leading sunspot was shown to appear closer to the polarity inversion line between the sunspots, than the following one. This result and the conclusion that, in the bulk of the investigated pairs of the magnetically conjugate sunspots, $\alpha_{min-L}<\alpha_{min-F}$ are closely coupled.
 
\end{abstract}
\end{opening}
%-------------------------------------------------

\section{Introduction}
     \label{S-Introduction} 
Sunspot umbra magnetic properties are discussed in many studies (see monographs by \citep{Bray1964,Obridko1985}, reviews \citep{Solanki2003,Borrero2011}, Ph.D. thesis \citep{Joshi2014}, papers\citep{Keppens1996,Jin2006,Otsuji2015,Tiwari2015,Tlatov2015,Zhivanovich2016} and references therein). Most often, studied were the properties of single sunspots, or, individually, properties of leading or following sunspots. Already in early papers, revealed was a positive correlation between the sunspot umbra magnetic field value and the umbra area. The minimal field line inclination angle to the vertical (to the positive normal toward the Sun surface) was established to be relatively small (close to zero) \citep{Bray1964,Keppens1996}.

In our previous studies \citep{Zagainova2015,Zagainova2017}, we studied in detail the umbra magnetic properties of magnetically-conjugate (leading/following) sunspots, i.e., the opposite-field polarity sunspots connected by magnetic field lines. For the first time, the field line minimal inclination angle to the vertical in umbrae of leading sunspots, $\alpha_{min-L}$, was shown to be smaller, than $\alpha_{min-F}$, the field line minimal inclination angle in the following spot ($\alpha_{min-L}<\alpha_{min-F}$), in the bulk of such sunspots. We assume that the $\alpha_{min}$ angle features the inclination angle of the magnetic tube connecting the leading and the following spots to the axis vertical. In other words, in the bulk of the magnetically-conjugate sunspot pairs, the axis of the magnetic tube connecting them is closer to the vertical direction in leading sunspots, as compared with following spots, which results in the origin of asymmetry of the magnetic tube connecting two types of sunspots. In \citep{Zagainova2015,Zagainova2017}, we also obtained the dependences of the magnetic induction maximum separately in umbrae of the leading and following magnetically-conjugate sunspots on the umbra area of these sunspots, $B_{max-L}(S_L)$ and $B_{max-F}(S_F)$. For the magnetically-conjugate sunspots, for which the $\alpha_{min-L}<alpha_{min-F}$ condition was met, the $B_{max}$ value was established, on average, to grow with the growth in $S$ and with the decrease in $\alpha_{min}$ for both types of sunspots. In \citep{Tlatov2015}, obtained were the dependences of the magnetic field on the umbra area separately for all the following and for all the leading sunspots over several cycles of solar activity. Note that the inverse correlation between the field value and the field line inclination angle, both in the umbra and in the penumbra, was revealed earlier, when analyzing these characteristics at different heights in the photosphere: with the height counted off from the photosphere base, the magnetic field modulus, on average, grows, whereas the magnetic field inclination angle decreases (see, for example, \citep{WestendorpPlaza2001,Tiwari2015} and references therein).

Note that, in the present-day theoretical sunspot models, the axis of the magnetic tube leaving an umbra is supposed to be vertical (i.e. $\alpha_{min}=0$; see, for example, \citep{Solovev2014}). But how much do such model ideas about properties of the umbra magnetic field agree with real features of the magnetic field in the umbrae of the observed sunspots?

The information on the umbra field characteristics, including the magnetic field inclination $\alpha$, is possible to obtain from their measurements by several ground-based and space-based instruments, involving the Hinode (SOT/SP) and SDO/HMI with high spatial resolution. At the same time, we note that the angle $\alpha$ is always determined by using measured characteristics of the magnetic field with an error, $\Delta\alpha$. By our estimates, the error of determining the sunspot umbra magnetic field inclination (by using the field vector measurements with the HMI instrument) varies within several tenths of degree to several degrees. Within such a scatter relative to the measured value $\alpha_{min-mes}(\alpha_{min-mes}\pm\Delta\alpha)$, there is a true value $\alpha_{min-true}$ in each spot. Here, $\alpha_{min-mes}$ is the $\alpha_{min}$ value obtained by using the magnetic field characteristics measured by a magnetograph. In this case, in some cases that we addressed, the $\alpha_{min}=0$ does not fit the $\alpha_{min-mes}\pm\Delta\alpha$ interval. I.e., at least, in some spots, $\alpha_{min}$ it is not equal to zero. In many other spots, whether $\alpha_{min}=0$ or $\alpha_{min}\neq0$ is impossible to establish.

Theoretical calculations \citep{Fan1993,Caligari1995}, according to which the leading and the following sunspots originate due to emergence of a curved magnetic tube from the convection zone depth, show that, in this case, the magnetic tube ''foot'' from the following sunspot umbra is more vertical, than the ''foot'' from the leading sunspot. The reason for that is the Coriolis force effect on the magnetic tube moving radially. In the above studies, the authors emphasized that the drawn conclusion agrees with the results in \citep{VanDrielGesztelyi1990}. Those results were based on analyzing the observations, according to which, in the bulk of the cases that the authors addressed, the polarity inversion line (PIL) between the adjacent ''hills'' of the photospheric field with the opposite polarity was closer to the ''hill'', whose polarity corresponded to the following sunspot. Therefore, \citep{VanDrielGesztelyi1990} considered the fact that the ''foot'' of the magnetic tube from the following sunspot umbra is more vertical, than the ''foot'' from the sunspot to be characteristic of magnetically-conjugate sunspot pairs. Note that a similar conclusion was later drawn in \citep{Cauzzi1998}, where the vector magnetograms acquired with Hawaii Stokes Polarimeter at Mees Solar Observatory, from October 1991 to June 1995. In other words, in the adjacent leading and following sunspots, the $\alpha_{min-L}>\alpha_{min-F}$ condition should be met more often, which does not agree with our results.

In this study, we investigated the PIL position relative to the sunspots of two types, and the association of distances from the sunspot center to PIL with the relation between $\alpha_{min-L}$ and $\alpha_{min-F}$ for a group of magnetically-conjugate sunspot pairs. Besides, we analyzed the umbra magnetic properties for the leading and following sunspots, both for sunspot pairs meeting the $\alpha_{min-L}<\alpha_{min-F}$ condition and the $\alpha_{min-L}>\alpha_{min-F}$ condition.
 
\section{Data and research methods} %%%%%%%%%%%%%%%%%%%%%%%%%%%%%%%%%%%%%%%%
      \label{S-general}      
We analyzed 74 pairs of the magnetically-conjugate sunspots observed over 2010 - 2015, which is by $70\%$ more, than in our previous studies \citep{Zagainova2015,Zagainova2017}. We use the term ''magnetically-conjugate sunspots'' for sunspot pairs with the magnetic field opposite polarity, whose umbrae are connected through field lines. The field lines were found by calculating the field in potential approximation with the use of an original program based on the results in \citep{Rudenko2001}. Also, we referred the cases, when separate field lines leaving the umbra of the same sunspot ended not precisely in the umbra of the other sunspot, but next to it, to magnetically-conjugate sunspots. Such cases may originate due to an insufficiently-high spatial resolution, with which the field calculations in potential approximation by using the potential field 90 spherical harmonic expansion were performed. The sunspot umbra magnetic field characteristics were found by using vector measurements of the photospheric field at high temporal and spatial resolutions of the Helioseismic and Magnetic Imager (HMI) instrument \citep{Schou2012} at the Solar Dynamics Observatory (SDO) \citep{Pesnell2012}. In this case, to obtain correct data for all the field components, we should solve the $\pi$-uncertainty problem for the field transverse component direction. In our study, this problem was resolved through a technique proposed in \citep{Rudenko2014}. This technique features an increased performance and precision of solution, as well as a possibility to use it at any distance from the solar disk center to the limb.

The angle $\alpha$ between the field direction and the radial direction from the Sun center was found by using the ratio: $cos(\alpha)=|B_r|/B$. Here, $B_r$ is the magnetic field radial component, and $B$ is the magnetic induction modulus. The $B_r$ value was found through the ratio including the measured values: $B$, angle $\delta$ between the field direction and the inclination, and the azimuth - angle $\psi$ measured in the sky plane counterclockwise between the CCD-matrix column direction and the field vector projection onto this plane. The sunspot positions were determined by the Sun images obtained with the SDO/HMI in continuum. To determine the sunspot umbra magnetic characteristics, we superposed the sunspots with the magnetic field distributions on the solar disk. Selected were the magnetically-conjugate sunspots in the groups observed beyond the periods of eruptive events on the Sun (flares, filament eruptions, coronal mass ejections). For each analyzed sunspot, we estimated the error for determining the angle $\alpha$ from the ratio: $\Delta\alpha=|[(d|B_r|/d\delta)\Delta\delta+(d|B_r|/d\psi)\Delta\psi]|/\sqrt{1-(|B_r|/B)^2}$. Here, $\Delta\delta$ and $\Delta\psi$ are the measurement errors $\delta$ and $\psi$ that are given together with the other magnetic field parameters measured with the HMI instrument $(B,\delta,\psi)$ for each pixel at each instant.

\section{Results} %%%%%%%%%%%%%%%%%%%%%%%%%%%%%%%%%%%%%%%%
      \label{R-general}
  \label{R1-labels}

 In this study, we corroborate the principal result of our previous studies: in the bulk of magnetically-conjugate sunspot pairs ($70\%$), $\alpha_{min-L}<\alpha_{min-F}$. We also show that there exists a statistically significant difference between the sample means of these values. This increases the robustness of the conclusion about the ratio of the minimal angles in leading and following sunspots.
 
Figure 1 shows an example of magnetically-conjugate sunspots (a), two magnetically - conjugate sunspots and PIL between them obtained through two techniques: at averaging $B_r$ through the running mean in the $[15\times15]$ pixel area (b) and in the $[50\times50]$ pixel area (c). For all the analyzed sunspot pairs, we found the distances along the lines connecting the umbra centers in the leading and following sunspots (from the sunspots to the intersection with PIL). First, determined were the distances from the leading sunspot umbra ($dL_l$) and from the following sunspot umbra ($dL_f$) to PIL. Further, determined were the parameters $dL=dL_l/dL_f$ and $AP_i=dL/(dL+1)$.

\begin{figure}    %%%%%%%%%%%%%%%%%% FIGURE 1
   \centerline{\includegraphics[width=1\textwidth,clip=]{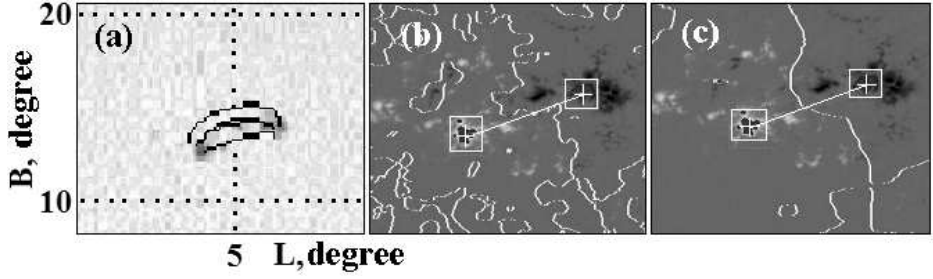}}
   \caption{Fig. 1. (a) example of a magnetically-conjugate leading and following sunspots. (b),(c) a photosphere fragment with the magnetic field radial component distribution, and the posed white-color isolines $B_r=0$. Shown are the field hills corresponding to the sunspots in figure (a), and the line connecting them that crosses PIL. (b) and (c) correspond to $B_r$ averaging over the $[15\times15]$ pixel and $[50\times50]$ pixel sites.}
   \label{Figure1}
\end{figure}

Our analysis shows that, unlike the conclusions in \citep{VanDrielGesztelyi1990}, in the bulk of the sunspot pairs that we investigated, $dL_l<dL_f$ (in $52.7\%$ cases for $[50\times50]$ pixel averaging and in $62.2\%$ cases of $[15\times15]$ pixel averaging). Herewith, for the sunspot pairs, for which this condition is met, the sunspots, for which $\alpha_{min-L}<\alpha_{min-F}$ are $66.7\%$ for the $[50\times50]$ pixel averaging and $73.9\%$ for the $[15\times15]$ pixel averaging. Figure 2(a)-(c) presents the $dL$ dependence on the relation between the inclination angles ($\alpha_{min-L}/\alpha_{min-F}$) of the magnetic tube connecting two types of sunspots for all the addressed sunspots, as well as for the sunspot groups with the $\beta$ configuration. From Fig. 2, it follows that the ratio between the distance from PIL to sunspots of two types is poorly associated with the ratio between the magnetic tube inclination angles to the radial direction from the Sun center.

\begin{figure}    %%%%%%%%%%%%%%%%%% FIGURE 2
   \centerline{\includegraphics[width=1\textwidth,clip=]{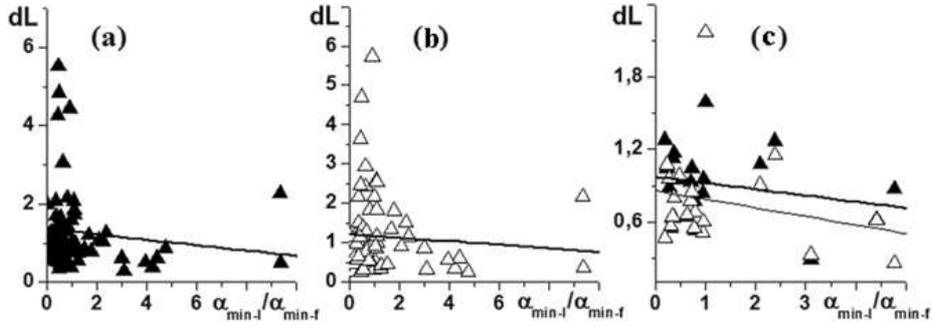}}
   \caption{$dL$ dependence on $\alpha_{min-L}/\alpha_{min-F}$. (a) $B_r$ averaging over $[50\times50]$ pixel sites; (b) $B_r$ averaging over $[15\times15]$ pixel sites; (c) the same as on panels (a), (b), but only for the sunspot groups with $\beta$-configurations, where the thick line shows averaging over $[50\times50]$ pixel sites, the thin line presents averaging over$[15\times15]$ pixel sites.}
   \label{Figure2}
\end{figure}

We built histograms for the $Ap_i$ parameter distribution (Fig. 3), like it was done in \citep{VanDrielGesztelyi1990}. That study showed that the mean value $<AP_i>=0.574$. In our case, $<AP_i>=0.503$ for the $50\times50]$ pixel averaging, and $0.47$ for the $[15\times15]$ pixel averaging. It is this obtained $<AP_i>$ value in \citep{VanDrielGesztelyi1990} that led to the conclusion that, in most cases, $\alpha_{min-L}>\alpha_{min-F}$. However, our analysis from the high spatial resolution data showed that there is no direct relation between the $<AP_i>$ values and the ratio between $\alpha_{min-L}$ and $\alpha_{min-F}$.

\begin{figure}    %%%%%%%%%%%%%%%%%% FIGURE 3
   \centerline{\includegraphics[width=1\textwidth,clip=]{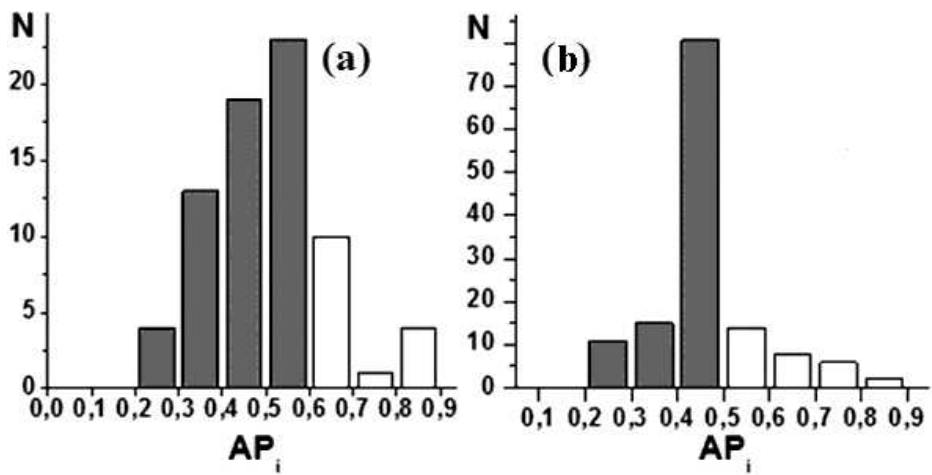}}
   \caption{Number histogram for the sunspot pairs depending on the $AP_i$ value, where (a) presents averaging over $[50\times50]$ pixel sites, (b) does the same for $[15\times15]$ pixel sites.}
   \label{Figure3}
\end{figure}

In our previous studies \citep{Zagainova2015,Zagainova2017}, we analyzed the umbra magnetic properties of leading and following sunspots only for the sunspot pairs, in which $\alpha_{min-L}<\alpha_{min-F}$ are met. In this paper, the sunspot magnetic properties were determined both for the sunspots satisfying the $\alpha_{min-L}<\alpha_{min-F}$, and the $\alpha_{min-L}>\alpha_{min-F}$ conditions. It appeared that such dependences like $\alpha_{min-L}(S_L)$, $\alpha_{min-F}(S_F)$, $B_{max-L}(S_L)$, and $B_{max-F}(S_F)$ feature identical trends, and are quantitatively close for the two addressed cases with a different asymmetry of the magnetic field  connecting them (Fig. 4). At the same time, the dependence of mean values of angles in umbra $<\alpha_{L,F}>$ on $S_{L,F}$ and on the mean value of the magnetic induction modulus $<B_{L,F}>$ appeared different for two sunspot groups (Fig. 5). In case of the pairs with $\alpha_{min-L}<\alpha_{min-F}$, there is practically no $<\alpha_{L,F}>$ dependence on $S_{L,F}$ and on $<B_{L,F}>$. For the sunspots meeting the $\alpha_{min-L}>\alpha_{min-F}$ condition, on average, the $<\alpha_{L,F}>$ value decreases with the growth in $S_{L,F}$ and in $<B_{L,F}>$.

\begin{figure}    %%%%%%%%%%%%%%%%%% FIGURE 4
   \centerline{\includegraphics[width=1\textwidth,clip=]{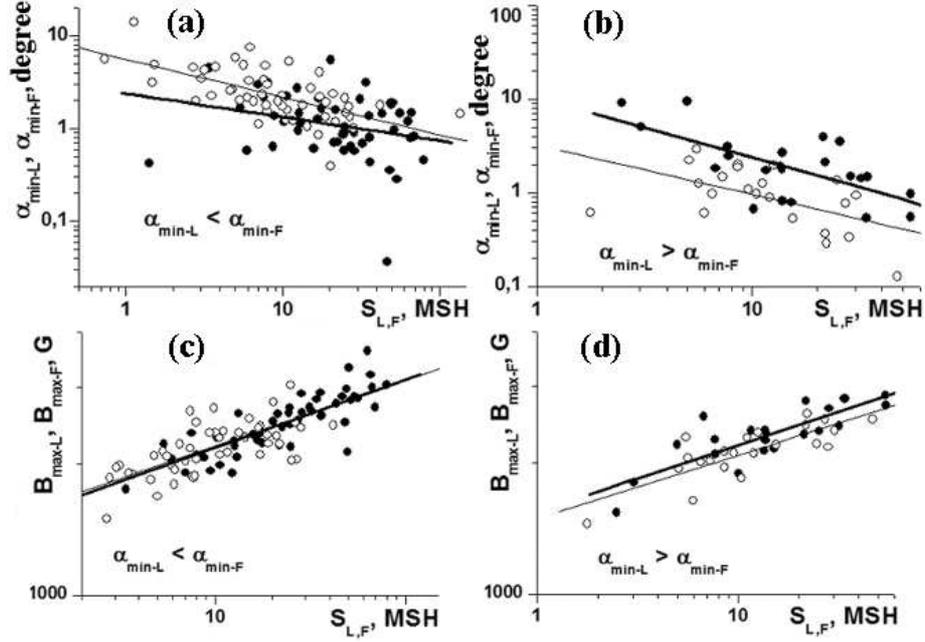}}
   \caption{Comparison between $\alpha_{min-L}(S_L)$ and $\alpha_{min-F}(S_F)$ in the sunspots meeting the $\alpha_{min-L}<\alpha_{min-F}$ condition (a) and the $\alpha_{min-L}>\alpha_{min-F}$ condition (b). Correspondingly, on panels (c) and (d) compared are the $B_{max-L}(S_L)$ and $B_{max-F}(S_F)$ dependences meeting the $\alpha_{min-L}<\alpha_{min-F}$ and $\alpha_{min-L}>\alpha_{min-F}$ conditions.}
   \label{Figure4}
\end{figure}

\begin{figure}    %%%%%%%%%%%%%%%%%% FIGURE 5
   \centerline{\includegraphics[width=1\textwidth,clip=]{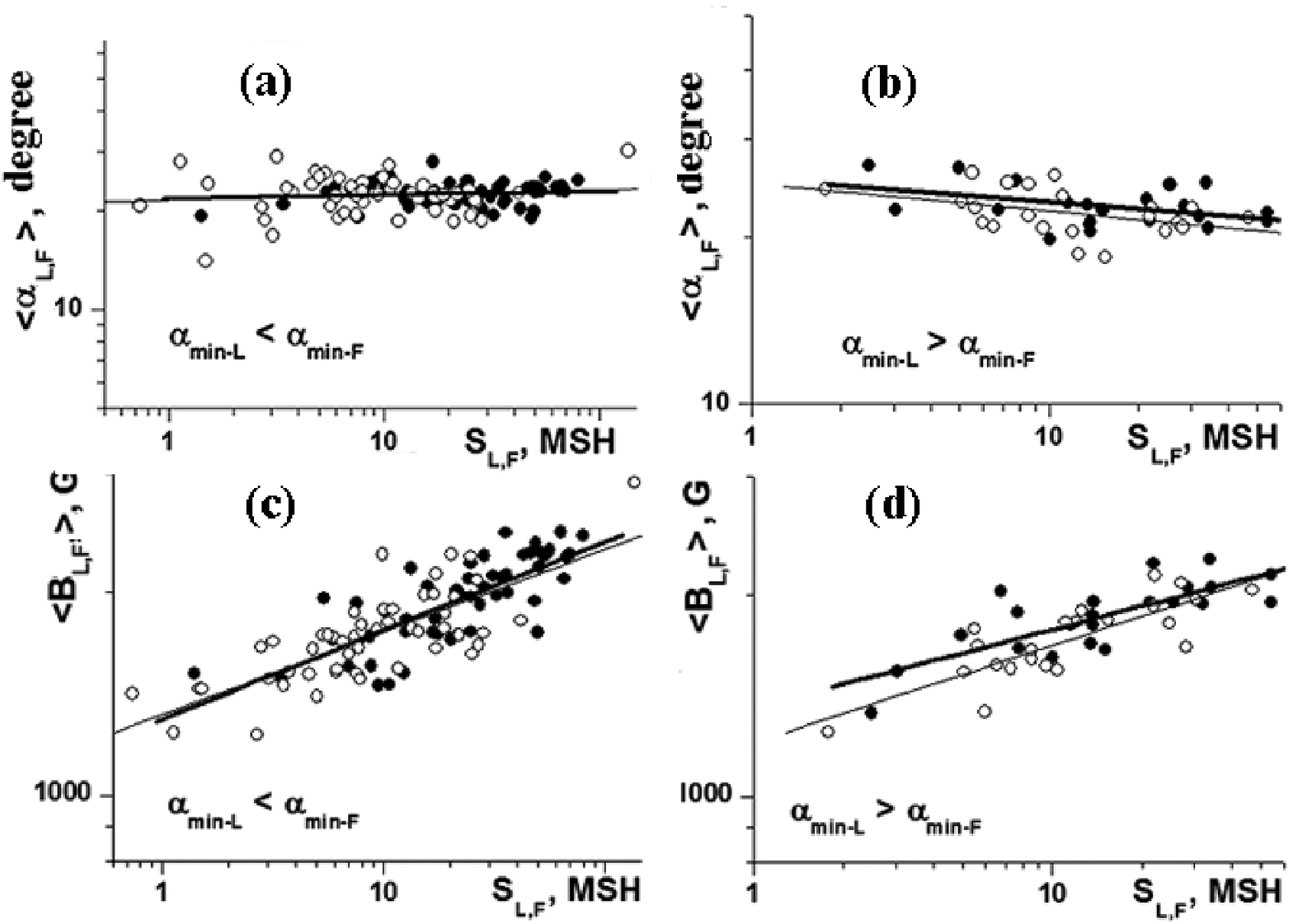}}
   \caption{Relation of the field line mean inclination angle $<\alpha_{L,F}>$ to the umbra area $S_{L,F}$ (a), and of the magnetic induction mean value $<B_{L,F}>$ (b) for the sunspots meeting the $\alpha_{min-L}<\alpha_{min-F}$ condition. Correspondingly, plots (c) and (d) show similar dependences for the sunspots meeting the $\alpha_{min-L}>\alpha_{min-F}$ condition.}
   \label{Figure5}
\end{figure}

Earlier in \citep{Zagainova2015,Zagainova2017}, a weak positive correlation between $\alpha_{min-L}$ and $\alpha_{min-F}$, as well as between $<\alpha_L>$ and $<\alpha_F>$ was shown to exist in the sunspots meeting the $\alpha_{min-L}<\alpha_{min-F}$ condition. In this paper, we revealed, whether there is a dependence between $B_{max-L}$ and $B_{max-F}$, as well as between $<B_L>$ and $<B_F>$, separately in the sunspots meeting both the $\alpha_{min-L}<\alpha_{min-F}$ and the $\alpha_{min-L}>\alpha_{min-F}$ conditions. The analysis showed that, for the two addressed cases, there is a positive correlation between the investigated parameters (Fig. 6). Herewith, for both sunspot groups, in some pairs of magnetically-conjugate sunspots, the $B_{max}$ and $<B>$ parameters accept great values in leading sunspots, as compared with those in the following ones, whereas in others, the situation is opposite. For the $\alpha_{min-L}>\alpha_{min-F}$ case, the $B_{max-F}(B_{max-L})$ and $<B_F>$ ($<B_L>$) dependences are flatter, than the similar dependences for the $\alpha_{min-L}<\alpha_{min-F}$ case. One may assume that such a behavior of angles and of the magnetic induction values in the umbrae of the investigated sunspots for the two addressed cases is because the magnetic tubes connecting the umbrae of the leading and of the following sunspots are more asymmetric for the $\alpha_{min-L}<\alpha_{min-F}$ case (i.e. the top of the field line connecting the two types of sunspots is more dramatically displaced toward the leading sunspot umbra, than it is for the $\alpha_{min-L}>\alpha_{min-F}$ case.

\begin{figure}    %%%%%%%%%%%%%%%%%% FIGURE 6
   \centerline{\includegraphics[width=1\textwidth,clip=]{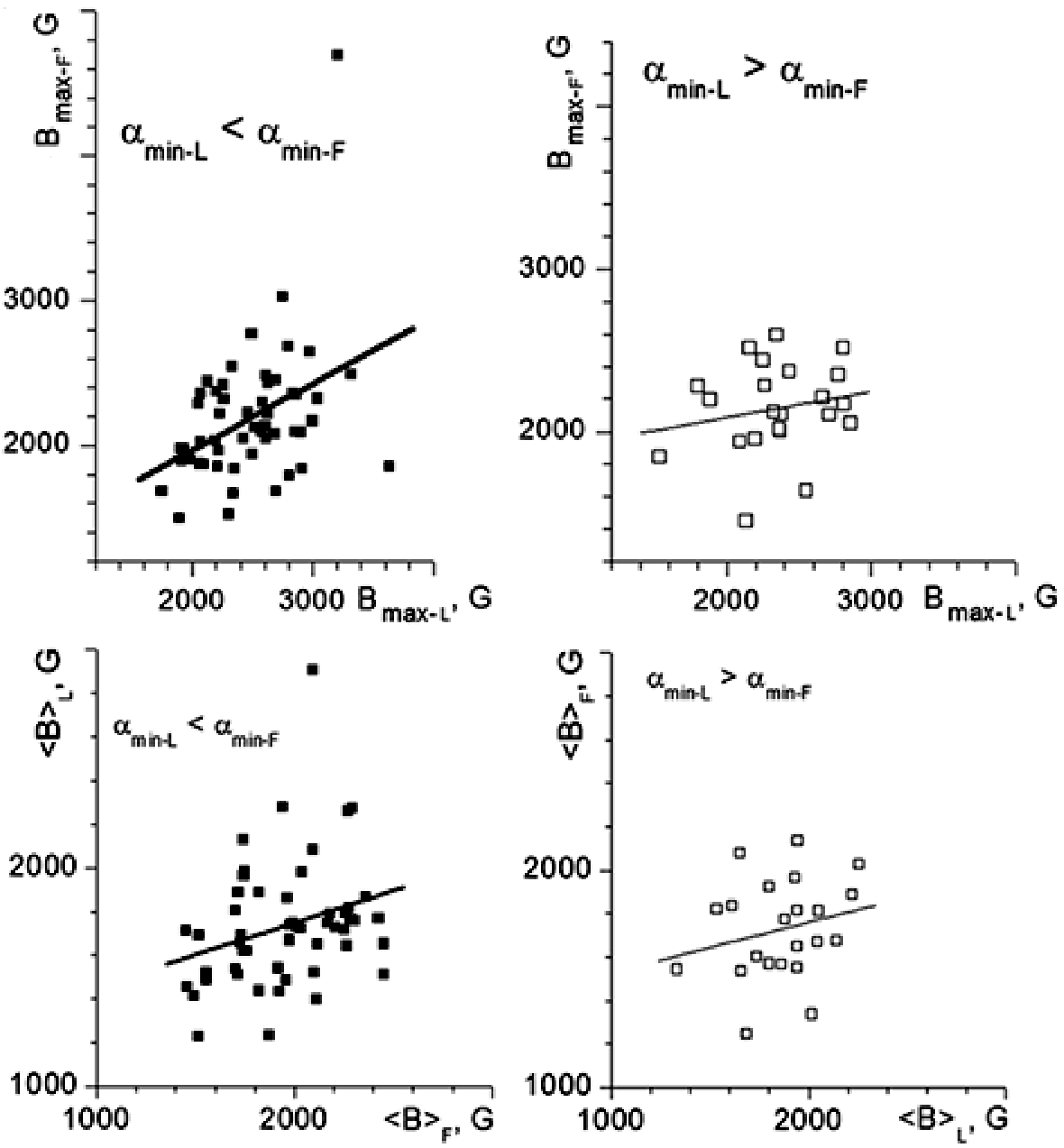}}
   \caption{Relation between $B_{max-L}$ and $B_{max-F}$ (a), and between $<B_L>$ and $<B_F>$ (b) for the sunspots meeting the $\alpha_{min-L}<\alpha_{min-F}$ condition. Correspondingly, panels (c) and (d) show similar dependences for the sunspots meeting the $\alpha_{min-L}>\alpha_{min-F}$ condition.}
   \label{Figure6}
\end{figure}

\section{Discussion and conclusions} %%%%%%%%%%%%%%
  \label{DC-labels}
  One of the main conclusions of this study is that the result obtained in our previous investigations \citep{Zagainova2015,Zagainova2017} has been corroborated for a new sample of magnetically-conjugate leading and following sunspots: the minimal inclination angle of the magnetic field in the umbra of the leading sunspot is smaller than that in the umbra of the following one ($\alpha_{min-L}<\alpha_{min-F}$) in the bulk ($70\%$ in this sample) of the investigated magnetically-conjugate sunspot pairs. We assume that $\alpha_{min-L,F}$ is the inclination angle of the axis of the magnetic tube connecting the leading and following sunspots, because the site, where the field inclination angle becomes minimal, practically coincides with the site, where the magnetic induction modulus becomes maximal.
  
The conclusion that the inclination angle of the magnetic tube connecting the leading and following sunspots may differ was drawn earlier by \citep{VanDrielGesztelyi1990} from the analysis of the PIL position relative to the adjacent field hills with the opposite polarity, and, later, it was corroborated based on theoretical calculations of the emergence (from the convection zone depth) of the magnetic tube, whose feet are thought to form magnetically-conjugate sunspots in the solar atmosphere \citep{Fan1993,Caligari1995}. In all those studies, the authors arrived at the conclusion that the magnetic tube connecting the leading and the following sunspots is more vertical in the umbra of the following sunspot, than that in the umbra of the leading one. This does not agree with our result, according to which the magnetic tube is more vertical in the leading sunspot, than that in the following one for the bulk of the addressed magnetically-conjugate sunspot pairs.

To reveal the reason for the contradiction between our results and the conclusions in \citep{VanDrielGesztelyi1990}, we investigated the PIL position relative to the magnetically-conjugate leading and following sunspots, by using vector measurements of the magnetic field with the HMI instrument. Unlike the conclusions in \citep{VanDrielGesztelyi1990}, where the photospheric field PIL appeared to be closer to the field hill with the polarity corresponding to the following sunspot, in the bulk of the pairs that we investigated, PIL between the leading and the following sunspots is closer to the leading sunspot. We assume that the difference in the results of two studies concerning the PIL proximity to the leading or to the following sunspot is caused by different properties of the samples used for the analysis, and by different quality of the used magnetic measurements. Van Driel-Gesztelyi and Petrovay (1990) analyzed the PIL position relative to the magnetic field hills with the opposite polarity. It is possible that these hills were not always related to the sunspots, and even if they were related to sunspots, then those sunspots were not always magnetically-conjugate, although they had the opposite field polarity. The precision of detecting PIL impacts what type of sunspots PIL is closer to. In our investigations, this was governed by the character of averaging the field radial component distributions on the solar disk. Unfortunately, there are no comments in \citep{VanDrielGesztelyi1990} on the PIL detection precision. Probably, it is the peculiarities of detecting PIL that caused the result by \citep{VanDrielGesztelyi1990} about the preferential PIL proximity to the following sunspots.

The contradiction between our results and the conclusions by \citep{Fan1993,Caligari1995} may be resolved, if one assumes that the formation of magnetically - conjugate sunspots occurs not due to emergence of a magnetic tube (with a solid cross-section) from the convection zone depth, but otherwise. Parker (1979) assumed that a sunspot may form from a number of thin magnetic flux tubes with a larger magnetic field that, in the sunspot region, merge in a uniform magnetic structure; at a depth of several hundreds of km to the photosphere base, they split into individual magnetic tubes. Weak plasma flows are supposed to exist between these tubes, and those flows produce a cross-section compression. But Parker did not explain, where and how originate those thin tubes, of which an umbra forms. 

Observations show that the formation of a bipolar sunspot pair occurs in a much more complex manner, than this could occur due to the emergence of a solid cross-section magnetic flux tube from under the photosphere. Qualitatively, the process of forming a magnetically-conjugate sunspot pair (with reference to supporting observations) is described in \citep{Solovev2014}. A new magnetic flux emergence precedes the sunspot origin. At the initial stage of the new magnetic flux emergence, the sunspot formation region looks like ''a complex mishmash of magnetic elements of different polarity (it is the region of the appearance of the top of a large-scale $\Omega$-shaped magnetic flux loop split into a lot of fine filaments tangled by convection)'' on magnetograms. Gradually, there occurs separation of the field polarities and formation of a typical bipolar structure.

Also, the shallow sunspot model developed recently does not agree with the idea of a bipolar sunspot pair formation due to the emergence of a curved magnetic flux tube from the convection zone bottom (see \citep{Solovev2014}). Fig. 2 in \citep{Solovev2014} illustrates such a sunspot formation. The idea of a shallow sunspot model originated due to advances in local helioseismology (LHelioS) (see \citep{Kosovichev2006,Kosovichev2009,Kosovichev2012} and references therein). A sharp increase in the plasma temperature (approximately by $1000^{\circ}$ as compared with the surroundings) was shown to occur in the location of sunspots at the 4 Mm depth. In other words, under the region of cold substance and strong magnetic field in a sunspot, originates a hot region with an abrupt transition from cold medium to the hot one. According to \citep{Kosovichev2012}, ''recent observations and radiative MHD numerical models\ldots lead to the understanding of sunspots as self-organized magnetic structures in the turbulent plasma of the upper convection zone, which are maintained by a large-scale dynamics\ldots''.

Thus, the magnetically-conjugate sunspot pair formation, most likely, may not be caused by an emergence (or by an emergence only) of a curved magnetic tube from the convection zone bottom. This is a more complex process involving processes at relatively small depths under the photosphere.

Below, we endeavor to answer the question: how robust is our conclusion of a relation between the field inclination angles in leading and following sunspots? First of all, we note that the magnetic tube axis is considered vertical in, practically, all sunspot models. It is supposed such even in models for magnetically-conjugate sunspots (see, for example, Fig. 3 in \citep{Solovev2014}). Sometimes, disregarding the features of measurements with magnetic field magnetographs, one concludes that, even in reality, the magnetic tube axis is vertical in all the observed sunspots. We note that there are no grounds to draw such a conclusion by using the measurements of the magnetic field characteristics even with the best up-to-date instruments. First of all, this is related to that the magnetic field inclination angle is always found with an error. As noted above, by our estimates, the precision of determining the inclination angle with the HMI instrument ranges from several tenths of a subdegree to several degrees. This means that, as long as the measured value of the umbra field minimal inclination angle is $\alpha_{min-mes}$, and the $\alpha$-angle measurement error in the site, where $\alpha=\alpha_{min}$, is $\Delta\alpha$, then the true value of $\alpha_{min}(\alpha_{min-true}$ is within: $\alpha_{min-true}=\alpha_{min-mes}\pm\Delta\alpha$. In some cases, the angle $\alpha_{min}=0^{\circ}$ fits the indicated range, but there are cases, when $\Delta\alpha$ was less, than the measured value $\alpha_{min-mes}$, and then $\alpha_{min}=0^{\circ}$ does not fit the $\alpha_{min-mes}\pm\Delta\alpha$ range. This does not give grounds to consider that, at least in some sunspots, the magnetic flux tubes leaving their umbrae are not vertical.

In some cases, for example, when finding the magnetic field properties in sunspots from the Hinode data, very small values of the field inclination angles near $0^{\circ}$ (and including $0^{\circ}$) are, practically, impossible to measure precisely due to the procedure used for the field measurements (see, for example, \citep{Otsuji2015}). This is related to that the magnetic field recovery is done by solving a Stokes polarization parameter inverse problem. Because the inverse problem solution is unsteady, the field values ''jump'' (vary dramatically) from point to point, i.e., one observes great and small values of the field horizontal component close to each other, ''alternately''. Hence, both small ($0^{\circ}$) and large ($180^{\circ}$) angle values are the artifacts emerging when solving the inverse problem, and they are not at all related to the real distribution of the magnetic field in an active region.

That the magnetic tube leaving an umbra may not be vertical was already shown in \citep{Kuklin1985} to explain the Wilson effect. Kuklin \citep{Kuklin1985} showed that the Wilson effect can be elucidated by assuming that the angle between the umbra plane and the normal to the Sun surface is different from $90^{\circ}$. In this case, the angle between the magnetic tube axis from the umbra and the normal to the Sun surface is not $0^{\circ}$, even if the magnetic tube axis inclination angle to the normal to the umbra plane is equal to $0^{\circ}$.

We adduce another argument in favor of that the sunspot umbra magnetic tube inclination angle to the normal to the Sun surface in many sunspots is not equal to zero. Like it was shown in our previous studies \citep{Zagainova2015,Zagainova2017}, on average, the umbra area and the magnetic field maximal value in leading sunspots are more than those in following sunspots. This means that the magnetic flux leaving the umbra of a leading sunspot may not entirely hit upon the following sunspot. Its part leaves aside either to other sites of the active region, where the analyzed sunspots are, or in other active regions. Our calculations (in potential approximation) for the field lines leaving the umbra of the leading sunspot corroborate this. In this case, the axis of the magnetic tube banding connecting the leading and the following sunspots may be displaced relative to the umbra geometrical center in the leading sunspot, which may lead to its departure from the vertical.

And another consideration that does not enable to think that the axis of the magnetic tube from a sunspot umbra should be vertical. As long as, in all the simultaneously observed sunspots, the axes of the magnetic tubes leaving the sunspot umbrae are oriented vertically, we should assume that the layer, within which the line used to measure the magnetic field forms, is radius-constrained by spherical surfaces. Most likely, this is a strong idealization, and this layer is subject to warping depending on latitude and longitude. This will lead to that the magnetic tube from the umbra, from the field vector measurements, does not appear vertical.

In our previous studies \citep{Zagainova2015,Zagainova2017}, we compared the umbra magnetic properties of leading and following sunspots only for the spots that met the $\alpha_{min-L}<\alpha_{min-F}$ condition. In this paper, we compared the umbra magnetic field characteristics of leading and following sunspots, and also established a relation of these characteristics to the umbra area in each type of sunspots for the sunspot pairs with a different asymmetry type of the field lines connecting them, i.e., both for the sunspots meeting the $\alpha_{min-L}<\alpha_{min-F}$ and the $\alpha_{min-L}>\alpha_{min-F}$ conditions. The $\alpha_{min-L}(S_L)$, $\alpha_{min-F}(S_F)$, $B_{max-L}(S_L)$, and $B_{max-F}(S_F)$ dependences are shown to feature identical trends, and they are quantitatively close for two sunspot groups with a different asymmetry of the field connecting them. A positive correlation was revealed to exist between $B_{max-L}$ and $B_{max-F}$, and, also, between $<B_L>$ and $<B_F>$, separately in the sunspots meeting the $\alpha_{min-L}<\alpha_{min-F}$ and the $\alpha_{min-L}>\alpha_{min-F}$ conditions. Herewith, the relation between these parameters in the sunspot groups with $\alpha_{min-L}<\alpha_{min-F}$ is stronger, than that in the group with $\alpha_{min-L}>\alpha_{min-F}$.

We also established that there is a positive correlation between the field maximal values, $B_{max-L}$ and $B_{max-F}$, as well as between the mean values, $<B_L>$ and $<B_F>$, in leading and following magnetically - conjugate sunspots. 

The authors thank the SDO and SOLIS teams for a possibility to freely use their data. This study was supported by the Russian Foundation for Basic Reserarch, grants No. 14-02-00308 and No. 15-02-01077. V.G. Fainshtein and G.V. Rudenko also did this study within the ISTP SB RAS R\&D Plan for 2016-2018, II.16.1.6. ''Geoeffective processes in the chromosphere and in the corona of the Sun'' (base project).
  
%\section*{Bibliography included manually }
%%% BIBLIOGRAPHY %%%%%%%%%%%%%%%%%%%%%%%%%%%%%%%%%%%%%%%%%%%%%%%%%%%%%%%%%%%

\end{article} 


\begin{thebibliography}{}
\bibitem[\protect\citeauthoryear{{Borrero et al.}}{2011}]{Borrero2011}
Borrero,~J.M., Ichimoto,~K.: 2011, \textit{Living Reviews in Solar Physics} \textbf{8}, Issue 1, article id. 4, 98.

\bibitem[\protect\citeauthoryear{{Bray et al.}}{1964}]{Bray1964}
Bray,~R., Loughhead,~R.: 1964, Sunspots (International Astrophysics Series). V. 7. London: Chapman and Hall.

\bibitem[\protect\citeauthoryear{{Caligari et al.}}{1995}]{Caligari1995}
Caligari,~P., Moreno-Insertis,~F., Schussler,~M.: 1995, \apj{} \textbf{441}, 886.

\bibitem[\protect\citeauthoryear{{Cauzzi and van Driel-Gesztelyi}}{1998}]{Cauzzi1998}
Cauzzi,~G., van Driel-Gesztelyi,~L.: 1998, Asymmetric Magnetic Field Distribution in Active Regions, Synoptic Solar Physics, \textit{ASP Conference Series}, ed. by K. S. Balasubramaniam, J.W. Harvey, and D. Rabin. V. 140, 105.

\bibitem[\protect\citeauthoryear{{Jin et al.}}{2006}]{Jin2006}
Jin,~C.L., Qu,~Z.Q., Xu,~C.L., Jhang,~X.Y., Sun,~M.G.: 2006, \apss{} \textbf{306}, 23. http://dx.doi.org/10.1007/s10509-006-9217-6.

\bibitem[\protect\citeauthoryear{{Joshi}}{2014}]{Joshi2014}
Joshi,~J.: 2014, Magnetic and Velocity Field of Sunspots in the Photosphere and Upper
Chromosphere. PhD Thesis, Technische Universit\"at Braunschweig, ISBN 978-3-944072-01-2, uni-edition GmbH 2014.

\bibitem[\protect\citeauthoryear{{Fan et al.}}{1993}]{Fan1993}
Fan,~Y., Fisher,~G.H., DeLuca,~E.E.: 1993, \apj{} \textbf{405}, 398.

\bibitem[\protect\citeauthoryear{{Keppens et al.}}{1996}]{Keppens1996}
Keppens,~R., Martinez Pillet,~V.: 1996, \aap{} \textbf{316}, 229.

\bibitem[\protect\citeauthoryear{{Kosovichev}}{2006}]{Kosovichev2006}
Kosovichev,~A.G.: 2006, \adv \textbf{38}, 876. 

\bibitem[\protect\citeauthoryear{{Kosovichev}}{2009}]{Kosovichev2009}
Kosovichev,~A.G.: 2009, \ssr{} \textbf{144}, 175.

\bibitem[\protect\citeauthoryear{{Kosovichev}}{2012}]{Kosovichev2012}
Kosovichev,~A.G.: 2012, \solphys{} \textbf{279}, 323.

\bibitem[\protect\citeauthoryear{{Kuklin}}{1985}]{Kuklin1985}
Kuklin,~A.G.: 1985, \textit{Researches in geomagnetism, aeronomy and physics of the Sun} \textbf{73}, 52 (In Russian).

\bibitem[\protect\citeauthoryear{{Obridko}}{1985}]{Obridko1985}
Obridko,~V.N.: 1985, Sunspots and complexes of activity, \textit{Moscow: Nauka}, 256 (in Russian).

\bibitem[\protect\citeauthoryear{{Otsuji et al.}}{2015}]{Otsuji2015}
Otsuji,~K., Sakurai,~T., Kuzanyan,~K.: 2015, \textit{Publ. Astron. Soc. Japan} \textbf{67}, 6.

\bibitem[\protect\citeauthoryear{{Pesnell et al.}}{2012}]{Pesnell2012}
Pesnell,~W.D., Thompson,~B.T., Chamberlin,~P.C.: 2012, \solphys{} \textbf{275}, 108, Issue 1-2, 3.

\bibitem[\protect\citeauthoryear{{Rudenko}}{2001}]{Rudenko2001}
Rudenko,~G.V.: 2001, \solphys{} \textbf{198}, Issue 1, 5.

\bibitem[\protect\citeauthoryear{{Rudenko et al.}}{2014}]{Rudenko2014}
Rudenko,~G.V., Anfinogentov,~S.A.: 2014, \solphys{} \textbf{289}, Issue 5, 1499.

\bibitem[\protect\citeauthoryear{{Schou et al.}}{2012}]{Schou2012}
Schou,~J., Scherrer,~P.H., Bush,~R.I., Wachter,~R., Couvidat,~S., Rabello-Soares,~M.C., et~al.: 2012, \solphys{} \textbf{275}, 229.

\bibitem[\protect\citeauthoryear{{Solanki et al.}}{2003}]{Solanki2003}
Solanki,~S.K.: 2014, \aapr{} \textbf{11}, 153. 

\bibitem[\protect\citeauthoryear{{Solov'ev and Kirichek}}{2014}]{Solovev2014}
Solov'ev,~A., Kirichek,~E.: 2014, \textit{Astrophysics and Space Science} \textbf{352}, Issue 1, 23. 

\bibitem[\protect\citeauthoryear{{Tiwari et al.}}{2015}]{Tiwari2015}
Tiwari,~S.K, van Noort,~M., Solanki,~S.K., Lagg,~A.: 2015, \aap{} \textbf{583}, A119. 

\bibitem[\protect\citeauthoryear{{Tlatov et al.}}{2015}]{Tlatov2015}
Tlatov,~A.G., Tlatova,~K.A., Vasil'eva,~V.V., Pevtsov,~A.A., Mursula,~K.: 2015, \adv{} \textbf{55}, Issue 3, 835.

\bibitem[\protect\citeauthoryear{{Van Driel-Gesztelyi and Petrovay}}{1990}]{VanDrielGesztelyi1990}
Van Driel-Gesztelyi,~L., Petrovay,~K.: 1990, \solphys{} \textbf{126}, 285.

\bibitem[\protect\citeauthoryear{{Westendorp Plaza et al.}}{2001}]{WestendorpPlaza2001}
Westendorp Plaza,~C., del Toro Iniesta,~J.C., Ruiz Cobo,~B., Marti\'inez Pillet,~V., Lites,~B.W., Skumanich,~A.: 2001, \apj{} \textbf{547}, 1130.

\bibitem[\protect\citeauthoryear{{Zagainova et al.}}{2015}]{Zagainova2015}
Zagainova,~Yu.S., Fainshtein,~V.G., Obridko,~V.N.: 2015, \textit{Geomagn. Aeron.} \textbf{55}, no. 1, 13.

\bibitem[\protect\citeauthoryear{{Zagainova et al.}}{2017}]{Zagainova2017}
Zagainova,~Yu.S., Fainshtein,~V.G., Obridko,~V.N., Rudenko,~G.V.: 2017, \textit{Asronomy Rep.} \textbf{61}, no. 6, 533.

\bibitem[\protect\citeauthoryear{{Zhivanovich et al.}}{2016}]{Zhivanovich2016}
Zhivanovich,~I., Solov'ev,~A.A., Smirnova,~V.V., Riehokainen,~A., Nagnibeda,~V.G.: 2016, \textit{Astrophysics and Space Science} \textbf{361}, 102, 6.

\end{thebibliography}
\end{document}